\documentclass[a4paper]{article}

\usepackage{url} 
\usepackage{times} 
\usepackage{verbatim}
\usepackage[T1]{fontenc}
\usepackage[swedish,english]{babel}
\usepackage{amsmath}
\usepackage{amssymb}
\usepackage[dvipdf]{graphicx}
\usepackage{color}
\usepackage{url}
\usepackage[utf8]{inputenc}
\usepackage[braket, qm]{qcircuit}
\usepackage{tikz}
\usepackage{graphicx}
\usetikzlibrary{positioning}

\usepackage{epsfig}
\usepackage{float}
\usepackage[small]{caption}
\usepackage{subfig}
\usepackage{eurosym}
\usepackage{mathtools}
\graphicspath{{.}}

\def\logit{\mathop{\rm logit}\nolimits}

\def\bin{\mathop{\rm Bin}}
\def\R{\mathbb{R}}
\def\C{\mathbb{C}}

\def\F{\mathcal F}
\def\Hilb{\mathcal H}

\def\RY{\mathrm{R_Y}}
\def\RZ{\mathrm{R_Z}}

\let\norm\undefined 
\DeclarePairedDelimiter\norm{\lVert}{\rVert}

\title{Quantum support vector regression \\ for disability insurance}

\author{
	Boualem Djehiche\footnote{Department of Mathematics, KTH Royal Institute of Technology, Sweden, {boualem@kth.se}. }
	\and 
    Björn Löfdahl \footnote{Department of Mathematics, KTH Royal Institute of Technology, Sweden, {bjornlg@kth.se}. }
}

\date{\today}

\begin{document}
\maketitle

\begin{abstract}
We propose a hybrid classical-quantum approach for modeling transition probabilities in health and disability insurance. The modeling of logistic disability inception probabilities is formulated as a support vector regression problem. Using a quantum feature map, the data is mapped to quantum states belonging to a quantum feature space, where the associated kernel is determined by the inner product between the quantum states. This quantum kernel can be efficiently estimated on a quantum computer. We conduct experiments on the IBM Yorktown quantum computer, fitting the model to disability inception data from a Swedish insurance company.
\end{abstract}

\vspace{4mm}

\section{Introduction}

Support vector machines (SVM) were first introduced as part of Vapnik's Statistical Learning Framework \cite{Vapnik2013}. Support vector classification aims to classify data, e.g. to determine if a picture contains a cat or a dog, whereas support vector regression (SVR) is used to model real-valued quantities, such as mortality rates or financial asset returns. All SVM's exploit the so-called kernel trick, where an optimization problem with data that has been mapped into a high- or even infinite-dimensional feature space may be efficiently solved by considering its Wolfe-dual \cite{Scholkopf2000}, for which the necessary input is reduced to a so-called kernel matrix consisting of inner products in the feature space between all data pairs. For cases where the kernel matrix can be readily determined, the corresponding optimization problem can be efficiently solved.

Rebentrost {\it et. al.} \cite{Rebentrost2014} showed that an SVM can be implemented on a quantum computer. This work was recently expanded on by Schuld and Killoran \cite{Schuld2019} and Havlicek {\it et. al.} \cite{Havlicek2019}. In essence, two related methods have been proposed. The first method consists of encoding data in a high-dimensional quantum feature space, calculating a quantum kernel, and subsequently using a variational quantum circuit to find a separating hyperplane. A second approach proposes to use a quantum computer to estimate the kernel, and to implement the resulting SVM optimization on a classical computer, a so-called hybrid classical-quantum implementation. Quantum kernel methods can be efficiently used to solve some optimization problems where the kernel cannot efficiently be determined on a classical computer. This was recently demonstrated by Liu {\it et. al.} \cite{Liu2021}.

In the insurance literature, SVMs have been used as a mortality graduation technique \cite{Kostaki2011support}. They could equally well be used to model other transition probabilities, such as disability inception or termination rates. These quantities are often estimated using classical techniques such as maximum likelihood or splines \cite{aro2015stochastic, djehiche2018hidden, RenshawHaberman, RenshawHaberman2000, Christiansen}. In this paper, we propose a hybrid classical-quantum SVR model for logistic disability inception probabilities, using a quantum kernel that can be estimated on a quantum computer. We conduct experiments on the IBM Yorktown quantum computer using disability inception data from a Swedish insurance company.

This paper is organised as follows. In Sections \ref{sec:svr} and \ref{sec:quantum_kernel}, we review kernel theory, support vector regression and quantum kernel estimation. In Section \ref{sec:model}, we propose a support vector regression model with a quantum kernel for disability inception rates. In section \ref{sec:inception_swedish}, we estimate the kernel matrix associated with disability inception data from a Swedish insurance company on a quantum computer. This kernel is then used in a support vector regression to estimate disability inception rates. The results are compared to those from classical support vector regression.

\section{Kernels and support vector regression}\label{sec:svr}
In this section we review kernel theory and support vector regression. Closely following \cite{Schuld2019}, we let $x_i\in \R^d, i=1,\ldots,n,$ denote observations in a data set, and let the mapping $\Phi : \R^d \mapsto \F$ be a feature map that maps a sample data point $x$ to a feature vector $\Phi(x)$ in a (usually higher-dimensional) feature space $\F$, usually taken as a Hilbert space. The mapping $\Phi$ naturally gives rise to a so-called kernel through the relation
\begin{equation}\label{eq:kernel_simple}
K(x,z) = \langle\Phi(x),\Phi(z)\rangle,
\end{equation}
\noindent
where $\langle\cdot,\cdot\rangle$ denotes the inner product on $\mathcal{F}$. Note that, since $K(x,z)$ is determined by the inner product of $\Phi(x)$ and $\Phi(z)$, it can be seen as a similarity measure between $x$ and $z$ in the feature space. The reproducing kernel Hilbert space (RKHS) associated with $\Phi$ is defined by
\begin{equation}\label{eq:rkhs}
\mathcal{R} = \{ f:\R^d \mapsto \C;\,\, \ f(x) = \langle w,\Phi(x)\rangle \ \ \forall \ x \in \R^d, w \in \F\}.
\end{equation}
\noindent
Note that the functions $\langle w,\Phi(x)\rangle$ can be interpreted as linear models in the feature space $\F$. Now, assume that we are given a cost function $\mathcal{C}$ that measures the goodness of fit of a model by comparing predicted values $\{f(x_i)\}_i$ with observed values $\{y_i\}_i$, and that has a regularization term $g(\norm{f})$, where $g$ is a strictly increasing function. Then, any function $f\in\mathcal R$ that minimizes the cost function $\mathcal{C}$ can be written as
\begin{equation}\label{eq:cost_minimizer}
f(x)=\sum_{i=1}^n\alpha_iK(x, x_i),
\end{equation}
\noindent
for some parameters $\alpha_i\in \R,\ i=1,\ldots, n.$ 

Perhaps the most famous application of the kernel approach is support vector regression (SVR) \cite{Vapnik2013}. SVR can be formulated as a convex optimization problem of the form
\begin{align*}
\textrm{P:}\,\, \min_{w,b,\xi, \xi^\prime} \ & \frac{1}{2}\norm{w}^2 + C\sum_{i=1}^n(\xi_i+\xi_i^\prime)&\\
\textrm{s.t.} \ \ \ & (w^T\Phi(x_i) + b) - y_i \leq  \varepsilon - \xi_i,&\ i=1,\ldots,n,\\
&  y_i - (w^T\Phi(x_i) + b) \leq  \varepsilon - \xi_i^\prime,&\ i=1,\ldots,n,\\
& \xi_i, \xi_i^\prime \geq 0,&\ i=1,\ldots,n,
\end{align*}
\noindent
where $\varepsilon$ determines the error tolerance of the solution, $C$ is a regularization parameter, and $\xi_i\in \R$ and $\xi_i^\prime\in\R, i=1,\ldots, n$, are slack variables. It can be shown \cite{Scholkopf2000} that the dual formulation of P is given by

\begin{align*}
\textrm{D:}\,\underset{\lambda, \lambda^\prime}{\max} & -\frac{1}{2}\sum_{i,j=1}^n(\lambda_i-\lambda_i^\prime)(\lambda_j-\lambda_j^\prime)K(x_i,x_j)- \varepsilon \sum_{i=1}^n(\lambda_i-\lambda_i^\prime) +  \sum_{i=1}^ny_i(\lambda_i-\lambda_i^\prime)&\\
\textrm{s.t.} \ \ \ & \sum_{i=1}^n(\lambda_i-\lambda_i^\prime) = 0,&\\
&0 \leq \lambda_i \leq C, i=1,\ldots,n,\\
&0 \leq \lambda_i^\prime \leq C, i=1,\ldots,n,
\end{align*}
\noindent
and that the solutions of P and D coincide and are given by 
\begin{equation}\label{eq:PD_solution}
f(x)=\sum_{i=1}^n\alpha_iK(x, x_i) + \beta,
\end{equation}
\noindent
where $\alpha_i = \lambda_i - \lambda_i^\prime$. In order to fit the model \eqref{eq:PD_solution} to data, we must first determine the kernel matrix $K = \{K_{ij}\}, i,j = 1,\ldots,n,$ where $K_{ij} = K(x_i, x_j)$. In the classical paradigm, we would choose a tractable kernel such as the kernel corresponding to radial basis functions (the so-called Gaussian kernel), evaluate the kernel matrix, and finally fit the model to data by solving the optimization problem D. An alternative to classical kernels is provided by the so-called quantum kernels, which we will briefly review in the following section.

\section{Quantum kernel estimation}\label{sec:quantum_kernel}
In quantum kernel estimation, the kernel is determined by a quantum feature map. Following \cite{Rebentrost2014}, \cite{Schuld2019} and \cite{Havlicek2019}, we let $\Phi:x \mapsto \Phi(x)$ (or $\ket{\Phi(x)}$ using Dirac's notation) denote a quantum feature map that maps a data point to a quantum state which is an element $\Phi(x)$ of a Hilbert space $\Hilb$. Any quantum state $\psi\in\Hilb$ naturally satisfies the famous Schr\"odinger equation
\begin{equation}\label{eq:schrodinger}
 i\hslash\frac{\partial}{\partial t}\psi(t,x)=H\psi(t,x),\quad \psi(0,\cdot)\in \mathcal{H}\,\,\text{is given},
\end{equation}
\noindent
where $H$ is the Hamiltonian operator associated to the quantum system. If $H$ is time-independent, the solution to \eqref{eq:schrodinger} is given by
\begin{equation}\label{eq:schrodinger_sol}
\psi(t,x) = U(t)\psi(0,x),
\end{equation}

\noindent 
where the operator $U$ defined by 
\begin{equation}
U(t) = e^{-iH t/\hslash}
\end{equation}
\noindent 
is the unitary time evolution operator associated with $H$. Thus, in analogy with \eqref{eq:schrodinger_sol}, using the characterization of reproducing kernel Hilbert spaces (RKHS), it can be shown (see \cite{scholkopf2001generalized},\cite{Schuld2019}, and the references therein) that for every pair $(\Phi,x)$  there is an operator $U_{\Phi}(x)$, known in the field of quantum computing as a {\it feature embedding circuit}, that is implicitly determined by the relation
\begin{equation}\label{eq:unitary}
\Phi(x)= U_{\Phi}(x)\Omega_0,
\end{equation}
\noindent
where $\Omega_0$ (also denoted $\ket{0\ldots 0}$ using Dirac's notation) denotes the ground state i.e. the quantum state with the lowest energy level (associated with the smallest eigenvalue of the generator of the operator $U_{\Phi}(x)$). Further, let the kernel $K$ corresponding to $\Phi$ be given by
\begin{equation}\label{eq:kernel}
K(x,z) = |\langle\Phi(x),\Phi(z)\rangle|^2.
\end{equation}
\noindent
As mentioned above, $K(x,z)$ it is essentially a similarity measure between $x$ and $z$ in the quantum feature space $\mathcal{H}$. It should be noted that the definition of the kernel \eqref{eq:kernel} deviates from the form \eqref{eq:kernel_simple} that is common in the classical literature, in that it involves taking the absolute value squared of the inner product. This is due to the following: Using \eqref{eq:unitary}, the kernel can be written as
\begin{equation}\label{eq:kernel_estimate}
K(x,z) = |\langle\Phi(x),\Phi(z)\rangle|^2 = |\Omega^{\dagger}_0 \,U^\dagger_{\Phi}(z) U_{\Phi}(x)\,\Omega_0|^2,
\end{equation}
\noindent
that is, $K(x,z)$ is given by the probability of obtaining the measurement outcome $\Omega_0$ 
when measuring the quantum state $\Psi(x,z)$
defined by

\begin{equation}
\Psi(x,z)= U^\dagger_{\Phi}(z) U_{\Phi}(x)\Omega_0,
\end{equation}

\noindent
where $U^\dagger$ denotes the adjoint operator of $U$. The probability \eqref{eq:kernel_estimate} can be estimated on a quantum computer by loading the state $\Psi(x,z)$ 
into a quantum circuit. This circuit is then run multiple times, and \eqref{eq:kernel_estimate} is estimated by the frequency of $\Omega_0$-measurements. 
Hence, the form of the kernel \eqref{eq:kernel} allows us to readily estimate it using a quantum computer. The advantage of the quantum approach is that there exist kernels which are hard to evaluate classically that can, in theory, be efficiently determined by a quantum computer. This was recently demonstrated in \cite{Liu2021}.

\section{Model description}\label{sec:model}
We consider a population of insured individuals, divided into subgroups based on some common characteristics. Let $E_i$ be the number of healthy individuals from the population subgroup $i,\ i=1,\ldots, n,$ in a given disability insurance scheme. We denote by $D_i$ the number of  individuals falling ill amongst the $E_i$ insured healthy individuals. For each population subgroup $i$ there is some associated data $x_i\in \R^d$ which may e.g. contain information about age, gender, and other characteristics of the population subgroup at hand. We assume that the conditional distribution of $D_i$ given $E_i$ is binomial:
\begin{equation}\label{eq:bin0}
D_i\sim\bin(E_i,p(x_i)),
\end{equation}
where $p(x_i)$ is the probability that an individual randomly selected from $E_i$ falls ill. We propose to model the logistic disability inception probabilities using support vector regression:
\begin{equation}\label{eq:logit0}
\logit p(x) = \log\frac{p(x)}{1-p(x)}=\sum_{i=1}^n\alpha_iK(x, x_i) + \beta,
\end{equation}
\noindent
where $K \in \R^{n\times n}$ is a kernel matrix associated with the data $\{x_i\}_i,\ $ and $\beta \in \R$ and $\alpha_i\in \R,\ i=1,\ldots, n,$ are parameters to be estimated from historical data. We propose to fit the model using a weighted support vector regression, with the weight for each sample proportional to the population subgroup size $E_i$, placing higher importance on large subgroups where the sampling errors of the observed inception probabilities are lower. The logistic transform guarantees that the probabilities estimated from the model lie in their natural interval $(0,1)$.

In the classical paradigm we would choose a tractable kernel such as the kernel corresponding to radial basis functions or the linear kernel. We propose instead that this kernel be calculated using a quantum feature map, to be evaluated on a quantum computer. In order to fit the model \eqref{eq:logit0} on a quantum computer, we must first choose a specific quantum feature map. This, in turn, determines the layout of our quantum circuit through \eqref{eq:unitary}. There are many ways to choose a suitable feature map, see e.g.  \cite{Schuld2019}. We suggest to choose $\Phi$ such that it captures the richness of the data $x$ while still being simple enough to be run on today's limited and noisy quantum computers. For simplicity, we will now assume that our data is two-dimensional, i.e. $x_i \in \R^2$. Then, it is enough to use a two-qubit unitary operator to obtain estimates of $K$. To this end, we choose the unitary operator given by
\begin{equation}\label{eq:our_unitary}
U_{\Phi}(x_i) = \Big(I\otimes \RY(\pi x_{i,2})\Big)C_\RZ(\pi x_{i,2})\Big(\RY(\pi x_{i,2})\otimes \RY(\pi x_{i,1})\Big),
\end{equation}
\noindent
where $\mathrm{R_Y}$ and $\mathrm{R_Z}$ denotes rotations around the $\mathrm{Y}$ and $\mathrm{Z}$ axes of the Bloch sphere (a.k.a. the Riemann sphere), respectively, and $C_\RZ$ denotes a controlled $\RZ$ operation on the second qubit, using the first qubit as a control. A graphical representation of the quantum circuit that implements this unitary is presented as follows:

\scalebox{1.0}{
	\Qcircuit @C=1.0em @R=0.2em @!R { \\
		\nghost{ {q}_{0} :  } & \lstick{ {q}_{0} :  } & \gate{\mathrm{R_Y}\,(\pi x_{i,1})} & \ctrl{1} & \gate{\mathrm{R_Y}\,(\pi x_{i,2})} & \qw & \qw\\ 
		\nghost{ {q}_{1} :  } & \lstick{ {q}_{1} :  } & \gate{\mathrm{R_Y}\,(\pi x_{i,2})} & \gate{\mathrm{R_Z}\,(\pi x_{i,2})} & \qw & \qw & \qw\\ 
		\\ }}

To facilitate interpretation, we let $x_{i,1}$ be a dummy variable taking the value 1 if the population subgroup is male, and 0 otherwise, and $x_{i,2}$ be the associated age of the population subgroup, measured in centuries. First, we apply a $\mathrm{R_Y}\,(\pi x_{i,1})$-gate to $q_0$. This flips $q_0$ from $\ket{0}$ to $\ket{1}$ for male subgroups. Then, we perform a $\mathrm{R_Y}\,(\pi x_{i,2})$ rotation on $q_1$. This operation rotates the state of $q_1$ from $\ket{0}$ towards $\ket{1}$, with the angle of rotation increasing as the age of the subgroup increases. The $C_\RZ(\pi x_{i,2})$ performs an additional rotation around the $\mathrm{Z}$-axis of the Bloch sphere, with the angle of rotation increasing as the age of the subgroup increases. Note that this rotation is only performed if $q_0$ is in the state $\ket{1}$, i.e. if the subgroup is male. Finally, we perform a $\mathrm{R_Y}\,(\pi x_{i,2})$ rotation on  $q_0$. 

For each data pair $(x_i, x_j)$, we run this quantum circuit inserting the values of $x_i$, and then run the adjoint circuit inserting the values of $x_j$. Finally, we perform a measurement on the two qubits. This circuit is run multiple times, and $K(x_i, x_j)$ is estimated by the frequency of obtaining the measurement $\Omega_0\coloneqq\ket{00}$. 
The resulting quantum circuit can be graphically represented as

\scalebox{1.0}{
	\Qcircuit @C=1.0em @R=0.2em @!R { \\
		\nghost{ {q}_{0} :  } & \lstick{\ket{0}  } & \gate{\mathrm{R_Y}\,(\pi x_{i,1})} & \ctrl{1} & \gate{\mathrm{R_Y}\,(\pi x_{i,2})} & \qw & \qw\\ 
		\nghost{ {q}_{1} :  } & \lstick{ \ket{0}  } & \gate{\mathrm{R_Y}\,(\pi x_{i,2})} & \gate{\mathrm{R_Z}\,(\pi x_{i,2})} & \qw & \qw & \qw\\ 
		\nghost{c:} &  & \lstick{/_{_{2}}} \cw & \cw & \cw & \cw & \cw\\ 
		\\ }}

\scalebox{1.0}{
	\Qcircuit @C=1.0em @R=0.2em @!R { \\
		\nghost{ {q}_{0} :  } &  & \gate{\mathrm{R_Y}\,(-\pi x_{j,2})} & \ctrl{1} & \gate{\mathrm{R_Y}\,(-\pi x_{j,1})} & \meter & \qw & \qw & \\ 
		\nghost{ {q}_{1} :  } &  & \qw & \gate{\mathrm{R_Z}\,(-\pi x_{j,2})} & \gate{\mathrm{R_Y}\,(-\pi x_{j,2})} & \qw & \meter & \qw & \\ 
		\nghost{c:} & &  \cw & \cw & \cw & \dstick{_{_{0}}} \cw \cwx[-2] & \dstick{_{_{1}}} \cw \cwx[-1] & \cw & \\ 
		\\ }}
	
This circuit is designed to clearly separate male and female subgroups, and to gradually increase the dissimilarity between different age groups as the difference in ages increases. Note that for $i=j$, all rotations cancel out, and the circuit will measure the state $\ket{00}$ with probability 1, i.e. $K(x_i, x_i)= 1$ as expected.

\section{Numerical results}\label{sec:inception_swedish}
In this section, we estimate the kernel matrix associated with disability inception data from a Swedish insurance company. This kernel is then used in a support vector regression to estimate the logistic disability inception rates. The data consists of inception counts for 81 groups of individuals as well as the associated age and gender for each group.

\subsection{Estimating the kernel matrix}
\begin{figure}[!ht]
	\begin{center}
		\includegraphics[width=0.8\linewidth]{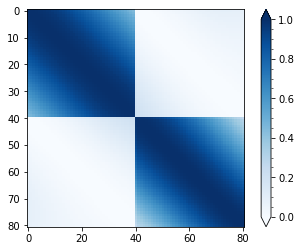}
		\caption{Kernel matrix determined by state vector simulation.}
		\label{fig:statevector_matrix}
	\end{center}
\end{figure}

We estimate the kernel matrix using the circuit from the previous section using two different techniques. Using \eqref{eq:kernel_estimate} and \eqref{eq:our_unitary}, we classically compute  $K(x_i,x_j)$ for each pair $(x_i, x_j)$ by matrix multiplication. Here, classically computing the kernel is possible due to the simple structure and low dimension of the unitary operator \eqref{eq:our_unitary}. This process is hereafter referred to as a state vector simulation. Figure \ref{fig:statevector_matrix} displays the estimated kernel matrix. This matrix has an interesting structure: it is block-diagonal. This is due to the fact that the second quadrant of the matrix correspond to the inner products of the female population groups. These share the common characteristic 'female', and each row is similar to its neighbours due to the encoding: similar ages are also similar in the quantum feature space. Analogously, the fourth quadrant of the matrix contains the male population groups. The first and third quadrants contain the inner products between male and female population groups, and so are dissimilar in the quantum feature space. 

\begin{figure}[!ht]
	\begin{center}
		\includegraphics[width=0.8\linewidth]{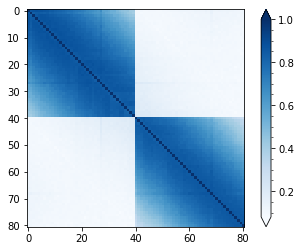}
		\caption{Kernel matrix estimated on the IBM Yorktown quantum computer.}
		\label{fig:yorktown_matrix}
	\end{center}
\end{figure}

Next, we run multiple experiments on the IBM Yorktown 5-qubit quantum computer (of which we use two qubits) to obtain an estimate of \eqref{eq:kernel_estimate}. For each data pair $(x_i, x_j)$, we run the circuit 8192 times, measure the outcomes, and estimate $K(x_i, x_j)$ with the observed frequency of the $\ket{00}$ state. Figure \ref{fig:yorktown_matrix} displays the estimated kernel matrix. Naturally, this simulation process introduces sampling error. Today's noisy and rather primitive quantum computers are unfortunately quite error prone, meaning that the total estimation error is often much larger than the sampling error. This issue can be partially mitigated using error correction techniques, see e.g. \cite{Temme2017}. We note that this matrix deviates somewhat from the kernel matrix obtained by state vector simulation, but it has the same characteristics of the block-diagonal structure and an increasing dissimilarity with increasing age difference. 

\subsection{Fitting Swedish disability inception rates}
We now fit the disability inception model \eqref{eq:logit0} to data with support vector regression, using four classical kernel methods, i.e. the linear kernel, a polynomial kernel of rank 3, the radial basis functions kernel, and a sigmoid kernel. In addition, we fit the model to data using the quantum kernels based on a state vector simulation as well as from the IBM Yorktown 5-qubit quantum computer. The models are fit using leave-one-out cross-validation, so that for each train-test-split, a single out-of-sample logistic disability inception rate is estimated. After applying the inverse logistic function to obtain a disability inception rate, we then calculate the weighted $R^2$ statistic for the out of sample rates, again using the population counts as weights. The results are presented in Table \ref{tab:R2}. The state vector quantum kernel performs better than three out of the four classical kernels, the exception being the polynomial kernel. The Yorktown quantum kernel is only slightly worse compared to the state vector simulation. 

\begin{figure}[htb]

\begin{minipage}{\textwidth}
\begin{tikzpicture}
  \node (img)  {\includegraphics[width=0.8\linewidth]{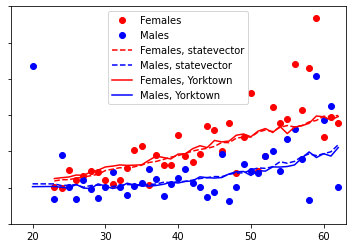}};
  \node[below=of img, node distance=0cm, yshift=1cm] {Age of population subgroup};
  \node[left=of img, node distance=0cm, rotate=90, anchor=center,yshift=-0.7cm] {Inception probability};
 
 \end{tikzpicture}
\end{minipage}%
\caption{Out-of-sample disability inception rates estimated by state vector simulation and from the IBM Yorktown quantum computer.}
\label{fig:inception_rates}
\end{figure}

The out-of-sample estimates are displayed in Figure \ref{fig:inception_rates}. Note that, due to confidentiality, the actual values of the estimates are not reported. The support vector regression approach manages to capture the difference between the genders, as well as finding a pattern in the age dimension. The middle-aged population groups are larger than the others, meaning that the highest weights will be placed on these ages for the purposes of calibrating the model. The observations with very high or very low ages are considered as outliers by the model, and so are virtually ignored. 

We note especially that the estimated inception rates from the Yorktown quantum kernel are comparable to the ones obtained from the state vector simulation, even though the estimated quantum kernels were themselves quite different. We believe that this is due to the fact that the characteristics of the kernel, namely the block-diagonal structure and the increasing dissimilarity with increasing age difference, were preserved, even though the actual kernel estimates differed significantly from each other. 

Seeing as the model under consideration fits the data well and produces errors that are comparable to today's classical methods, we conclude that estimating disability inception rates with quantum support vector regression is a viable statistical method even on today's noisy quantum computers. This bodes well for the future where complex and high-dimensional data might well be modeled and fitted accurately to data in a timely fashion using quantum computers.

        
\begin{table}[!ht]
	\caption{Weighted out-of-sample $R^2$ for the classical and quantum kernels.}
	\label{tab:R2}
	\footnotesize
	\begin{center}
		\begin{tabular}{c|cccc}
			
			kernel & $R^2$\\
			\hline 
			linear & 0.426\\
			polynomial & 0.550\\
			radial basis functions & 0.529\\
			sigmoid& 0.494\\
			state vector quantum kernel & 0.541\\
			Yorktown quantum kernel & 0.518\\
		\end{tabular} \\
	\end{center}
\end{table}

\newpage

\bibliography{qsvr_sjuk}
\bibliographystyle{acm}

\end{document}